\documentclass[twocolumn]{aastex701}

\usepackage{amsmath}

\begin{document}
	
\title{First Observational Evidence for QPO-like Coevolution between Characteristic Damping Timescales and X-ray Time Lags among AGNs}

\author[0000-0003-3392-320X]{Haoyang Zhang}
\affiliation{School of Physics and Electronic Engineering, Hanjiang Normal University, Shiyan 442000, China}
\affiliation{Key Laboratory of Astroparticle Physics of Yunnan Province, Yunnan University, Kunming 650091, China}
\email[show]{zhanghaoyang1@hjnu.edu.cn}  

\author[0000-0002-5880-8497]{Li Zhang}
\altaffiliation{Corresponding author}
\affiliation{Key Laboratory of Astroparticle Physics of Yunnan Province, Yunnan University, Kunming 650091, China}
\affiliation{Department of Astronomy,
Yunnan University,
Kunming 650091, China}
\email[show]{lizhang@ynu.edu.cn} 


\begin{abstract}
Quasi-periodic oscillations (QPOs) and stochastic variability provide complementary probes of the inner accretion flow around supermassive black holes in active galactic nuclei (AGNs). Previous multi-epoch studies of narrow-line Seyfert 1 galaxy RE~J1034+396 revealed a coevolution between the QPO frequency and the X-ray time lag, but whether the stochastic variability participates in the same structural evolution has remained unclear. We analyze the multi-epoch XMM-Newton observations of RE~J1034+396 and model the soft (0.3--1~keV) and hard (1--4~keV) light curves separately using a damped random walk process. We obtain reliable characteristic damping timescales (CDTs) for 17 observations, with the soft-band CDT consistently longer than its hard-band counterpart. When combined with the X-ray time lag, the hard-band CDT traces a counterclockwise closed loop that closely resembles the previously reported QPO-frequency--time-lag loop, whereas the soft-band CDT exhibits more complex trajectory. Under plausible dynamical, thermal, and viscous interpretations, the hard-band CDT is associated with characteristic scales in the inner hot accretion flow/corona. The observed loops may represent different projections of a common cyclic evolution of the inner accretion flow/corona, as the three timing observables may respond to its dynamical, stochastic, and radiative properties on different timescales. These results provide the first observational support among AGNs for the physical picture in which changes in the spatial extent of the hot inner flow/corona simultaneously affect QPO and stochastic variability.
\end{abstract}


\section{Introduction}\label{sec:intro}

Active galactic nuclei (AGNs) and black hole X-ray binaries (XRBs) are powered by accretion onto supermassive and stellar-mass black holes, respectively \citep{2006csxs.book..623T,2017A&ARv..25....2P}. Their X-ray emission can exhibit quasi-periodic oscillations (QPOs), which appear as narrow peaks above the stochastic continuum in the power spectral density (PSD) \citep{2006ARA&A..44...49R,2008Natur.455..369G}. QPOs may carry information about physical processes in the compact regions close to the black hole, including radiative mechanisms under strong gravity, the structural evolution of the corona or hot accretion flow, and the coupling between the compact inner region and the outer accretion disk (e.g., \citealt{2006ARA&A..44...49R,2014ARA&A..52..529Y,2025FrASS..1130392L}).

QPOs were first discovered in XRBs \citep{1979Natur.278..434S}, and recurrent QPOs have been reported in multi-epoch observations of several XRB sources, such as GRS 1915+105, XTE J1550--564, and GRO J1655--40 \citep{2006ARA&A..44...49R}. In XRBs, QPO frequency and X-ray time lag can evolve together. Such coevolution has been observed in GRS 1915+105 and several type-B QPO systems, where the time lag changes systematically with QPO evolution and can even reverse between hard- and soft-lag modes \citep{2000ApJ...541..883R,2017MNRAS.466..564G,2023MNRAS.525..854M}. Because most QPO candidates in AGNs have relatively low statistical significance or have been reported in only a single observing epoch \citep{2012A&A...544A..80G,2023ApJ...946...52Z}, the coevolution of QPO frequency with X-ray time lag has so far only been observed in narrow-line Seyfert 1 galaxy RE J1034+396 \citep{2024ApJ...961L..32X}, which shows a highly significant and recurrent QPO over multiple epochs \citep{2008Natur.455..369G,2014MNRAS.445L..16A,2020MNRAS.495.3538J,2023ApJ...946...52Z,2024ApJ...961L..32X}. This interesting coevolution may reflect a coupling between the QPO-driving mechanism and lag-producing processes in a structurally evolving corona or hot inner flow, including the propagation of accretion-rate fluctuations, the modulation of disk seed photons and coronal heating, and disk reprocessing \citep{2000ApJ...538L.137N,2021MNRAS.503.5522K,2025MNRAS.536.3284U,2025ApJ...989...59X}. However, the detailed physical origin of this coupling remains unclear.

In addition to the multi-epoch evolution of QPOs themselves, \citet{2011MNRAS.415.2323I} proposed that changes in the geometry of the corona or hot inner flow should also affect the properties of the PSD within the truncated-disk/hot-inner-flow framework. In this framework, the PSD bending frequency is associated with the viscous timescale at the outer edge of the hot inner flow, which can be expressed as a characteristic damping timescale (CDT; $\tau_{\rm damping}=(2\pi f_{\rm bend})^{-1}$). Since the QPO frequency is determined by the Lense--Thirring precession of the same hot inner flow, changes in the outer radius of the hot flow are expected to make the CDT evolve in a systematic pattern analogous to that of the QPO frequency. This model successfully explains the observed connection between spectral-state evolution and timing properties in XRBs (e.g., \citealt{2012MNRAS.419.2369I,2014MNRAS.440.2882R,2022MNRAS.509.2517C}). 

Multiple lines of evidence support a unified view in which AGNs and XRBs are governed by similar accretion physics across different black hole mass scales (e.g., \citealt{2006Natur.444..730M,2015ApJ...798L...5Z,2015SciA....1E0686S,2017SSRv..207....5R,2021MNRAS.500.2475J,2021Sci...373..789B,2024ApJ...967L..18Z,2025MNRAS.538.2161Z}). In this context, RE J1034+396 currently serves as a unique AGN probe for testing whether a dynamically evolving corona or hot inner flow can shape the timing properties of black hole accretion systems, because of its robust and recurrent multi-epoch QPO and the reported evolution of QPO frequency with X-ray time lag.

In this Letter, we analyze multi-epoch X-ray observations of RE J1034+396 and present the first observational evidence for QPO-like coevolution between CDTs and X-ray time lags among AGNs. This result provides the first support in an AGN for the physical picture that the dynamical evolution of the geometric scale of the corona or hot inner flow can shape the timing properties of black hole accretion systems. The rest of this Letter is structured as follows. We describe the observations and data reduction in Section~\ref{sec:d}. The data analysis methods and results are presented in Section~\ref{sec:d_r}. In Section~\ref{sec:dis}, we discuss the possible physical origins of the soft- and hard-band CDTs and explore how the main findings of this work provide new insights into the structure and evolution of the inner accretion flow in AGNs. Finally, we summarize our conclusions in Section~\ref{sec:conclusion}.

\begin{deluxetable*}{ccccccc}[t]
    \tablecaption{XMM-Newton Observation Log and Timing Measurements of RE J1034+396\label{tab:d_r}}
    \tablewidth{0pt}
    \setlength{\tabcolsep}{10pt}
    \tablehead{
        \colhead{Obs. ID} &
        \colhead{Date} &
        \colhead{Duration} &
        \colhead{$f_{\rm QPO}$} &
        \colhead{Time Lag} &
        \colhead{$\ln\tau_{\rm hard}$} &
        \colhead{$\ln\tau_{\rm soft}$} \\
        \colhead{} &
        \colhead{(UT)} &
        \colhead{(ks)} &
        \colhead{$\times10^{-4}$ Hz} &
        \colhead{(s)} &
        \colhead{(s)} &
        \colhead{(s)}   \\
        \colhead{(1)} &
        \colhead{(2)} &
        \colhead{(3)} &
        \colhead{(4)} &
        \colhead{(5)} &
        \colhead{(6)} &
        \colhead{(7)}
    }
    \startdata
    0506440101 (Obs-a)  & 2007-05-31 & 90 & $2.63 \pm 0.05$ & $-180 \pm 90$ & $6.06^{+0.47}_{-0.51}$ & $7.14^{+0.32}_{-0.28}$ \\
    0561580201          & 2009-05-31 & 55 & --              & --            & $7.90^{+1.17}_{-0.89}$ & $8.34^{+0.94}_{-0.86}$ \\
    0655310101          & 2010-05-09 & 44 & 2.70            & --            & $6.78^{+0.39}_{-0.33}$ & $7.89^{+0.68}_{-0.50}$ \\
    0655310201          & 2010-05-11 & 47 & 2.50            & --            & $6.35^{+0.30}_{-0.28}$ & $8.13^{+0.60}_{-0.42}$ \\
    0675440101          & 2011-05-27 & 33 & 2.60            & --            & $6.34^{+0.36}_{-0.33}$ & $7.32^{+0.53}_{-0.42}$ \\
    0675440201          & 2011-05-31 & 25 & 2.60            & --            & $6.26^{+0.41}_{-0.37}$ & $7.60^{+0.74}_{-0.52}$ \\
    0675440301          & 2011-05-07 & 27 & --              & --            &  --                    & --                     \\
    0824030101 (Obs-b)  & 2018-10-30 & 72 & $2.83 \pm 0.06$ & $430  \pm 50$ & $6.26^{+0.21}_{-0.19}$ & $7.32^{+0.28}_{-0.23}$ \\
    0865010101 (Obs-1)  & 2020-11-20 & 88 & $2.56 \pm 0.05$ & $107  \pm 87$ & $6.44^{+0.21}_{-0.19}$ & $8.83^{+0.50}_{-0.38}$ \\
    0865011001 (Obs-2)  & 2020-11-30 & 85 & $2.47 \pm 0.03$ & $-156 \pm 85$ & $6.53^{+0.22}_{-0.20}$ & $8.14^{+0.38}_{-0.30}$ \\
    0865011201 (Obs-3)  & 2020-12-02 & 91 & $2.57 \pm 0.05$ & $-429 \pm 87$ & $6.22^{+0.20}_{-0.19}$ & $7.38^{+0.27}_{-0.22}$ \\
    0865011101 (Obs-4)  & 2020-12-04 & 89 & $2.57 \pm 0.02$ & $-342 \pm 70$ & $6.29^{+0.21}_{-0.19}$ & $7.88^{+0.34}_{-0.26}$ \\
    0865011301 (Obs-5)  & 2021-04-24 & 92 & $2.83 \pm 0.05$ & $451  \pm 55$ & $6.37^{+0.19}_{-0.17}$ & $7.53^{+0.27}_{-0.22}$ \\
    0865011401 (Obs-6)  & 2021-05-02 & 87 & $2.74 \pm 0.05$ & $514  \pm 57$ & $6.52^{+0.20}_{-0.19}$ & $7.53^{+0.28}_{-0.24}$ \\
    0865011501 (Obs-7)  & 2021-05-08 & 91 & $2.60 \pm 0.05$ & $475  \pm 61$ & $6.63^{+0.21}_{-0.19}$ & $7.58^{+0.27}_{-0.22}$ \\
    0865011601 (Obs-8)  & 2021-05-12 & 89 & $2.61 \pm 0.03$ & $401  \pm 62$ & $6.46^{+0.19}_{-0.18}$ & $7.82^{+0.33}_{-0.26}$ \\
    0865011701 (Obs-9)  & 2021-05-16 & 92 & $2.63 \pm 0.06$ & $300  \pm 77$ & $6.65^{+0.20}_{-0.18}$ & $7.85^{+0.32}_{-0.25}$ \\
    0865011801 (Obs-10) & 2021-05-30 & 92 & $2.52 \pm 0.06$ & $-77  \pm 107$& $6.49^{+0.21}_{-0.19}$ & $7.82^{+0.33}_{-0.26}$
    \enddata
    \tablecomments{The columns are as follows: (1) the observation ID. The numbers in parentheses denote the observation numbering adopted by \citet{2024ApJ...961L..32X}, which is used throughout this work for convenient reference; (2) the observation date; (3) the EPIC-pn observation  duration; (4) and (5) the QPO frequency and the corresponding time lag, respectively. The QPO frequencies and time lags are taken from \citet{2008Natur.455..369G,2014MNRAS.445L..16A,2020MNRAS.495.3538J,2024ApJ...961L..32X}. Positive and negative values indicate hard- and soft-band lags, respectively. (6) and (7) the CDTs measured in the hard and soft energy bands, respectively. Dashes indicate that no statistically significant measurement was available for the corresponding quantity.}
\end{deluxetable*}

\section{Data Reduction} \label{sec:d}
RE J1034+396 has more than 20 independent observations in the XMM-Newton archive, spanning multiple epochs \citep{2001A&A...365L...1J}. We selected all observations with individual observation durations exceeding 10 ks (see columns (1)--(3) of Table~\ref{tab:d_r}). To remain consistent with the settings adopted in previous studies, we processed the EPIC-pn data using a time bin of 200 s and defined the energy ranges as 0.3--1 keV for the soft band and 1--4 keV for the hard band. Data reduction was performed using the official XMM-Newton Science Analysis System (SAS v19.0.1). 

The data reduction procedure was as follows: The EPIC-pn data were first reprocessed with the \texttt{epproc} task to generate calibrated event lists, which were combined when necessary. Good time intervals were then identified using \texttt{tabgtigen}. A background count-rate threshold of $\mathrm{RATE} \leq 0.4$ was applied to exclude intervals affected by enhanced particle background. The resulting good-time-interval files were subsequently used in \texttt{evselect} to produce cleaned pn event lists. Source events were extracted from circular regions centered on RE J1034+396. Radii of $40^{\prime\prime}$ and $50^{\prime\prime}$ were generally adopted for observations obtained in the small-window and large-window modes, respectively. For the large-window observations, background events were selected from an annular region with inner and outer radii of $50^{\prime\prime}$ and $100^{\prime\prime}$. For the small-window observations, a circular background region with a radius of $40^{\prime\prime}$ was chosen on the same CCD chip, sufficiently far from the source position. Before extracting the final light curves, possible photon pile-up was examined using \texttt{epatplot}. The source and background light curves were then extracted with \texttt{evselect}, and the background-subtracted light curves were produced using \texttt{epiclccorr}.
\section{Data Analyses and Results}\label{sec:d_r}

\begin{figure*}[t]
	\figurenum{1} \label{fig:exam}
	\gridline{\fig{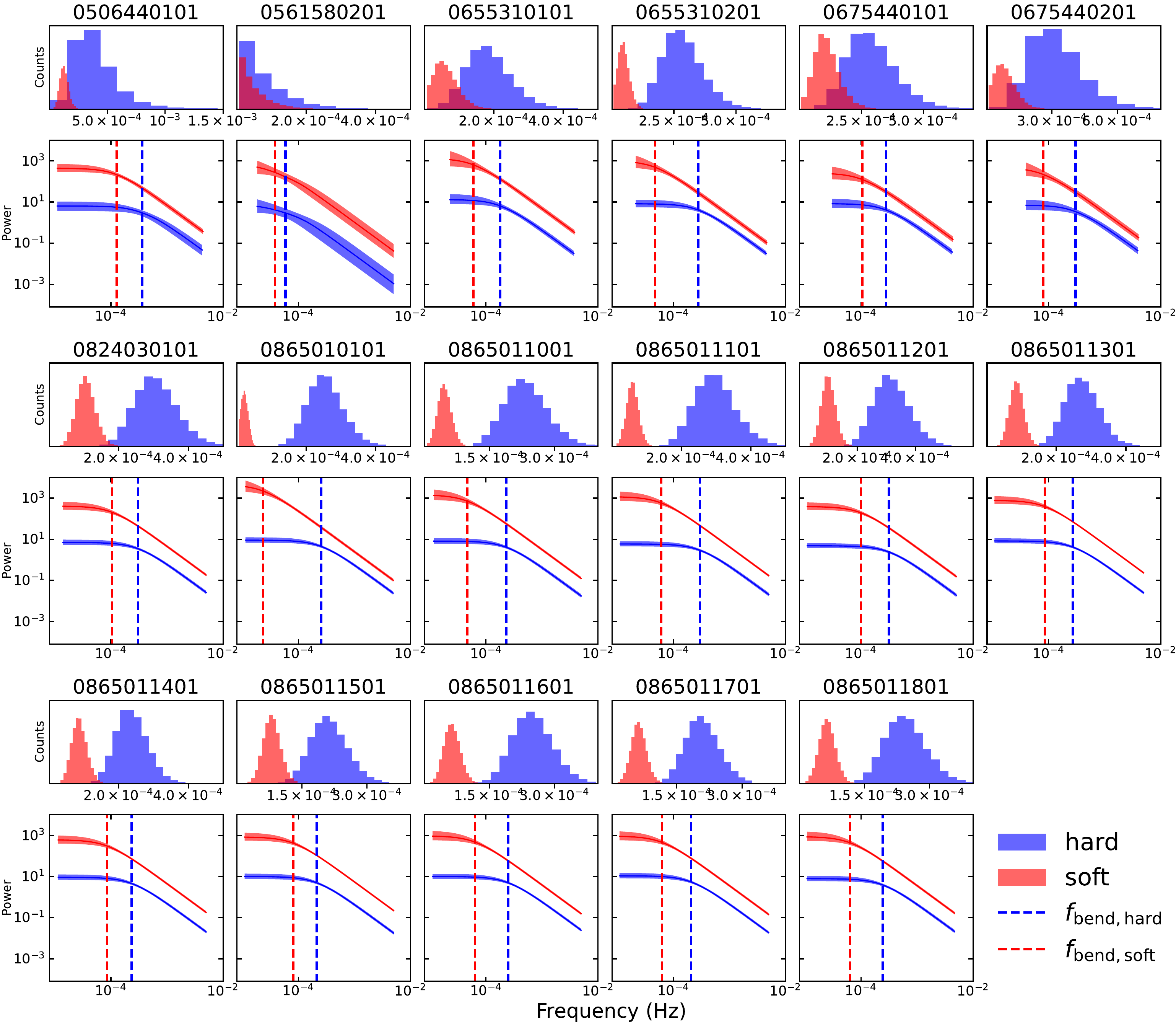}{\textwidth}{}
	}
	\caption{Posterior distributions of the CDTs and the corresponding DRW PSDs for the XMM-Newton observations of RE J1034+396. Each composite panel corresponds to one observation. The upper subpanels show the marginalized posterior distributions of the CDTs, while the lower subpanels display the PSDs calculated from the corresponding DRW models. Blue and red denote the hard and soft bands, respectively. The vertical dashed lines mark the bend frequencies inferred from the CDTs.\label{fig:psd}}
\end{figure*}

\subsection{The Damped Random Walk Model}
We focus on obtaining CDT from the multi-epoch observational data of this source. A large number of studies have shown that the multi-wavelength variability of AGN can be modeled as a damped random walk (DRW) process (e.g., \citealt{2009ApJ...698..895K,2010ApJ...721.1014M,2012ApJ...760...51R,2013ApJ...765..106Z,2017A&A...597A.128K,2021ApJ...907...96S,2021Sci...373..789B,2022ApJ...930..157Z,2024MNRAS.527.2672S,2024ApJ...967L..18Z,2025PhRvD.111h3049S,2026ApJ..1002..201X,2026ApJS..284...50C}). Specifically, from the perspective of observation, the DRW model can naturally take into account the measurement error of each observation data and directly obtain the PSD from the fitting of the time-domain signal without the need for Fourier transformation (the latter is affected by red noise leakage, aliasing, and the arbitrary selection of PSD bins). Theoretically, the energy injection and restoration to equilibrium in the accretion disk and jet of AGN can be naturally simplified into a DRW process \citep{2009ApJ...698..895K}.

The DRW model describes AGN variability as a stochastic process driven by random perturbations and damped toward a long-term mean on a characteristic timescale. In some literature, this process is also referred to as the first-order continuous-time autoregressive process \citep{2009ApJ...698..895K} or the Ornstein--Uhlenbeck process \citep{1996AmJPh..64..225G}. According to \citet{2014ApJ...788...33K}, the differential equation of this process can be expressed as

\begin{equation}
	\Bigg[\frac{d}{dt}+\frac{1}{\tau_{\text{DRW}}}\Bigg]\text{y}(t)=\sigma_{\text{DRW}}\epsilon(t),
\end{equation}
where $\tau_{\text{DRW}}$ is the CDT, $\sigma_{\text{DRW}}$ is the random perturbation amplitude, and $\epsilon(t)$ is the white noise. The PSD of a DRW model is

\begin{equation}
P(f)=\frac{2\sigma_{\rm DRW}^{2}\tau_{\rm DRW}^{2}}
{1+\left(2\pi\tau_{\rm DRW}f\right)^{2}}.
\end{equation}
Therefore, the bending frequency is $f_{\rm bend}=1/(2\pi\tau_{\rm DRW})$. 

In modeling, we utilized the software developed by \citet{2017AJ....154..220F}, which is capable of quickly performing Gaussian process fitting--\texttt{celerite}. The covariance function (kernal function) of a DRW process in \texttt{celerite} is defined as

\begin{equation}
	k(t_{nm})=2\sigma_{\text{DRW}}^{2}\cdot\text{exp}(-t_{nm}/\tau_{\text{DRW}}), \label{equ:kernal}
\end{equation}
where $t_{nm}$ is the time lag between two measurements. To determine the values of the parameters in Equation \ref{equ:kernal}, we first maximized the likelihood using the \texttt{L-BFGS-B} algorithm, and adopted the resulting estimates as the initial values for the subsequent Markov Chain Monte Carlo (MCMC) analysis. We then explored the posterior distributions of the model parameters using the \texttt{emcee} sampler \citep{2013PASP..125..306F}. A total of $10^{5}$ steps were generated, with the first 30000 steps discarded as burn-in. After obtaining the parameter estimates of the DRW model, the PSD of this model is

\begin{equation}
	P(\omega)=\sqrt{\frac{8}{\pi}}\sigma^{2}_{\text{DRW}}\tau_{\text{DRW}}\frac{1}{1+(\omega\tau_{\text{DRW}})^{2}}. \label{equ:model_psd}
\end{equation}
This formula can directly perform the Fourier transform on Equation \ref{equ:kernal} to obtain.

After modeling the AGN light curve, we assessed the goodness of fit. According to \citet{2017A&A...597A.128K,2021Sci...373..789B}, a reliable and physically meaningful CDT should satisfy two criteria: its upper limit should be shorter than one-tenth of the observational baseline, whereas its lower limit should exceed the mean cadence of the light curve. In addition, if the model adequately captures the red-noise variability of the AGN, the standardized residuals are expected to be consistent with white noise \citep{2014ApJ...788...33K}.

\subsection{Multi-epoch CDT Measurements}
We modeled each observation listed in Table~\ref{tab:d_r} with the DRW model using \texttt{celerite}, treating the soft and hard bands separately. The posterior medians and 1$\sigma$ confidence intervals of the CDTs derived from the MCMC analysis are presented in columns (6) and (7) of Table~\ref{tab:d_r}. With the exception of Obs. ID: 0675440301, all observations passed the parameter-reliability test. The Lomb--Scargle periodograms of the corresponding standardized residuals are generally consistent with white-noise behavior in logarithmic space (see Appendix~\ref{sec:appendix}). Because our model is designed to capture only the red-noise component of the time series, observations with highly significant QPO detections may retain peaks in the periodograms of their standardized residuals at frequencies close to the QPO frequencies identified in the original light curves.

Using Equation~\ref{equ:model_psd}, we present the PSDs derived from the modeling of each observation in Figure~\ref{fig:psd}. The blue and red lines and histograms represent the modeling results for the hard and soft energy bands, respectively. In each panel, the upper subpanel shows the posterior distribution of the CDT for the corresponding observation, while the lower subpanel displays its PSD. The dashed lines in the PSD panels mark the corresponding bend frequencies. In all 17 valid measurements, the soft-band CDT was consistently longer than its hard-band counterpart.

We note that the posterior distributions of the hard-band CDTs are generally broader than those of the soft-band CDTs. This difference may primarily arise from the lower hard-band count rate, which is, on average, approximately one order of magnitude lower than that in the soft band. The correspondingly larger Poisson noise and lower signal-to-noise ratio reduce the precision with which the characteristic turnover in the variability power spectrum can be constrained, leading to larger uncertainties in the inferred hard-band CDT. Nevertheless, nearly all of the hard-band posterior distributions remain unimodal and approximately Gaussian, indicating that the CDT measurements are reasonably well constrained despite their larger uncertainties.

\begin{figure*}[ht]
	\figurenum{2} 
	\gridline{\fig{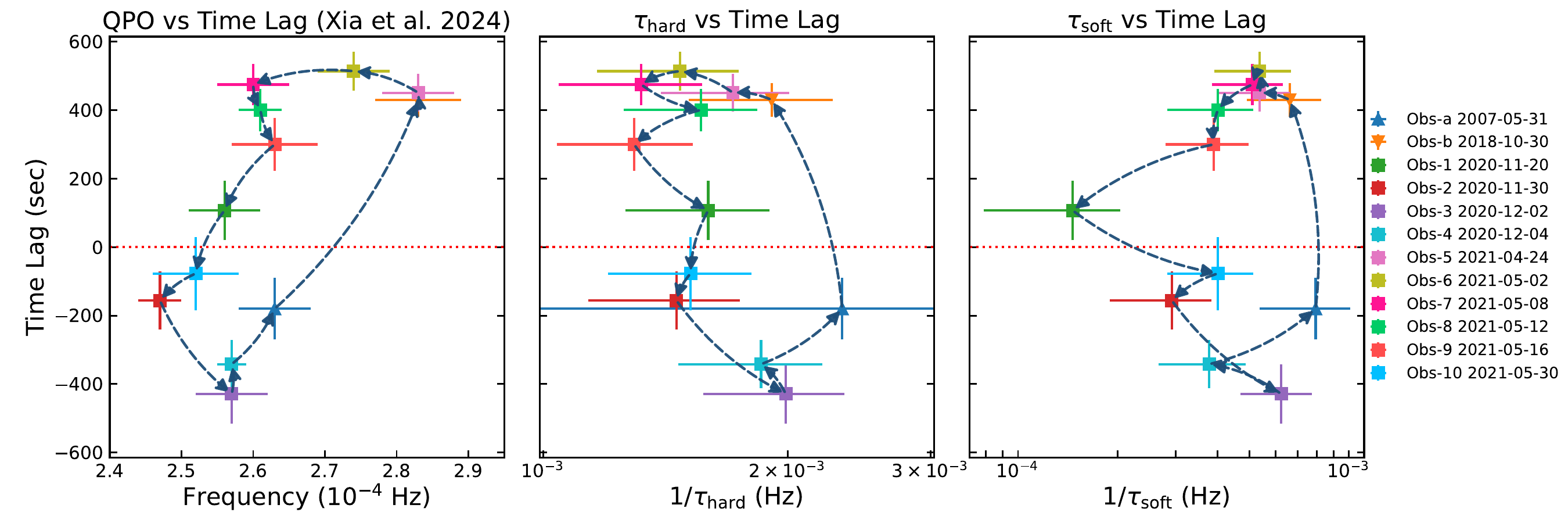}{\textwidth}{}
	}
	\caption{Comparison of the evolutionary loops formed by the X-ray time lag and different characteristic frequencies. The left panel shows the QPO frequency--time-lag loop reported by \citet{2024ApJ...961L..32X}, while the middle and right panels show the time lag as a function of the reciprocal hard- and soft-band CDTs, respectively. Different symbols and colors denote individual observations, and the error bars represent the corresponding measurement uncertainties. The dashed curves and arrows indicate the hypothetical evolutionary sequence proposed by \citet{2025ApJ...983...13X}, following Obs-b, Obs-5, Obs-6, Obs-7, Obs-8, Obs-9, Obs-1, Obs-10, Obs-2, Obs-3, Obs-4, and Obs-a. The red dotted horizontal line marks zero time lag. The hard-band CDT forms a loop that more closely resembles the QPO frequency--time-lag loop, whereas the soft-band CDT exhibits a more complex loop structure.\label{fig:loop}}
\end{figure*}

\subsection{QPO-like Coevolution between CDT and X-ray Time Lags}
Among these 17 valid measurements, previous studies have explored the coevolution between the QPO frequency and the time lag between the soft and hard energy bands \citep{2024ApJ...961L..32X}. Specifically, we focus on the densely sampled observations obtained between 2020 and 2021, corresponding to Obs-1--Obs-10 in column (1) of Table~\ref{tab:d_r}. Over this period, the QPO frequency and X-ray time lag trace an extremely rare closed loop in parameter space as a function of time, as shown in the bottom panel of Figure~4 in \citet{2024ApJ...961L..32X}, where positive values relative to the dashed reference lines indicate hard-band lags. By incorporating the earlier observations, \citet{2025ApJ...983...13X} proposed a hypothetical interconnected evolutionary cycle between the QPO frequency and X-ray time lag, in which the observations are arranged as Obs-b, Obs-5, Obs-6, Obs-7, Obs-8, Obs-9, Obs-1, Obs-10, Obs-2, Obs-3, Obs-4, and Obs-a, with the individual observations representing different phases of the cycle (see the left panel of Figure~\ref{fig:loop}). Interestingly, when the CDT measurements are combined with the X-ray time lag to define a parameter space, closed loops similar to this evolutionary sequence also emerge (see the middle and right panels of Figure~\ref{fig:loop}). By visual inspection, the loop formed by the hard-band CDT appears to more closely resemble the interconnected evolutionary cycle between the QPO frequency and X-ray time lag. In contrast, the soft-band loop exhibits a more complex internal structure.

Given the relatively large uncertainties of the measurements involved in the three loops, particularly those of the hard-band CDT, we further examined the robustness of their counterclockwise evolution using the signed loop area calculated with the shoelace formula (see Appendix~\ref{sec:appendix2}). We propagated the measurement uncertainties through $10^{5}$ Monte Carlo realizations by sampling each quantity according to its corresponding uncertainty distribution. The signed loop area was then calculated for each realization along the adopted evolutionary sequence. The resulting normalized signed loop areas are $3.28_{-0.93}^{+0.96}$, $4.67_{-3.71}^{+5.79}$, and $4.60_{-1.78}^{+2.07}$ for the QPO-frequency--lag, hard-band reciprocal-CDT--lag, and soft-band reciprocal-CDT--lag relations, respectively. The fractions of realizations with positive signed areas are 99.975\%, 90.506\%, and 99.746\%, respectively. Thus, all three relations favor the same counterclockwise evolution after the measurement uncertainties are taken into account, indicating that the observed loop-like patterns are generally robust.

\begin{figure*}[t]
	\figurenum{3} 
	\gridline{\fig{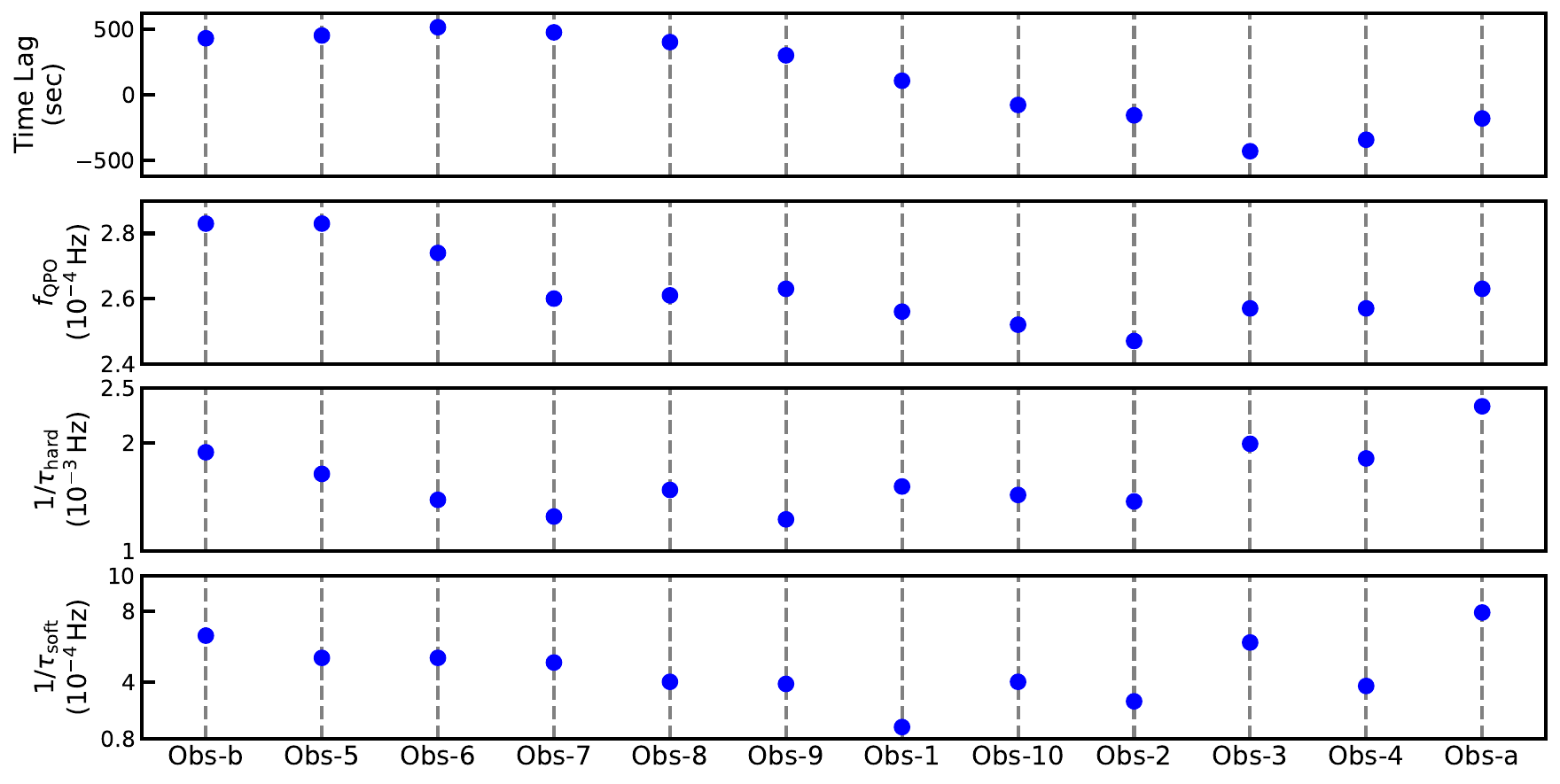}{\textwidth}{}
	}
	\caption{Evolution of the X-ray time lag, QPO frequency, and hard- and soft-band CDTs along the hypothetical evolutionary sequence proposed by \citet{2025ApJ...983...13X}. From top to bottom, the panels show the X-ray time lag, QPO frequency, $1/\tau_{\rm hard}$, and $1/\tau_{\rm soft}$, respectively. The observations are arranged as Obs-b, Obs-5, Obs-6, Obs-7, Obs-8, Obs-9, Obs-1, Obs-10, Obs-2, Obs-3, Obs-4, and Obs-a. The blue points represent the measurements for the individual observations, and the vertical dashed lines are included to guide the eye. The QPO frequency, X-ray time lag, and both hard- and soft-band CDTs exhibit systematic variations along the proposed evolutionary sequence, with relative phase offsets among them.\label{fig:sequence}}
\end{figure*}

In Figure~\ref{fig:sequence}, we further show the evolution of the measured quantities along the hypothetical evolutionary sequence. Along this sequence, the X-ray time lag remains at a high positive level from Obs-b to Obs-7 and reaches its maximum at Obs-6. It then decreases almost continuously, changes from a hard lag to a soft lag between Obs-1 and Obs-10, and reaches its minimum at Obs-3 before recovering toward Obs-a. The QPO frequency shows a broadly similar but earlier evolution: it is highest at Obs-b and Obs-5, decreases overall toward its minimum at Obs-2, and subsequently increases through Obs-3, Obs-4, and Obs-a. The reciprocal hard-band CDT decreases from Obs-b and reaches its minimum at Obs-9, after which it generally increases and attains its maximum at Obs-a. The reciprocal soft-band CDT follows a comparable overall trend, declining to its minimum at Obs-1 and then recovering toward its maximum at Obs-a, although several local fluctuations are present. Notably, the minima occur successively at Obs-9 for $1/\tau_{\rm hard}$, Obs-1 for $1/\tau_{\rm soft}$, Obs-2 for the QPO frequency, and Obs-3 for the X-ray time lag. On the opposite side of the cycle, both reciprocal CDTs peak at Obs-a, followed by the high QPO-frequency state at Obs-b--Obs-5 and the maximum time lag at Obs-6. This ordered displacement of the extrema indicates that the four timing quantities trace a common evolutionary pattern with distinct relative phase offsets. 

\begin{deluxetable}{ccc}
    \tablecaption{Radial Location Inferred from Different Timescale
    Prescriptions\label{tab:model}}
    \tablewidth{0pt}
    \setlength{\tabcolsep}{10pt}

    \tablehead{
        \colhead{Physical timescale} &
        \colhead{Model parameters} &
        \colhead{$R\ (r_{\rm g})$} \\
        \colhead{} &
        \colhead{($\alpha$, $H/R$)} &
    }

    \startdata
    Dynamical & --        & 4.01--27.56 \\ \hline
    Thermal   & 0.1, --   & 0.86--5.94 \\
              & 0.5, --   & 2.52--17.36 \\ \hline
    Viscous   & 0.1, 0.5  & 0.34--2.36 \\
              & 0.5, 0.5  & 1.00--6.89
    \enddata
\end{deluxetable}

\begin{figure}[ht]
	\figurenum{4}
	\gridline{\fig{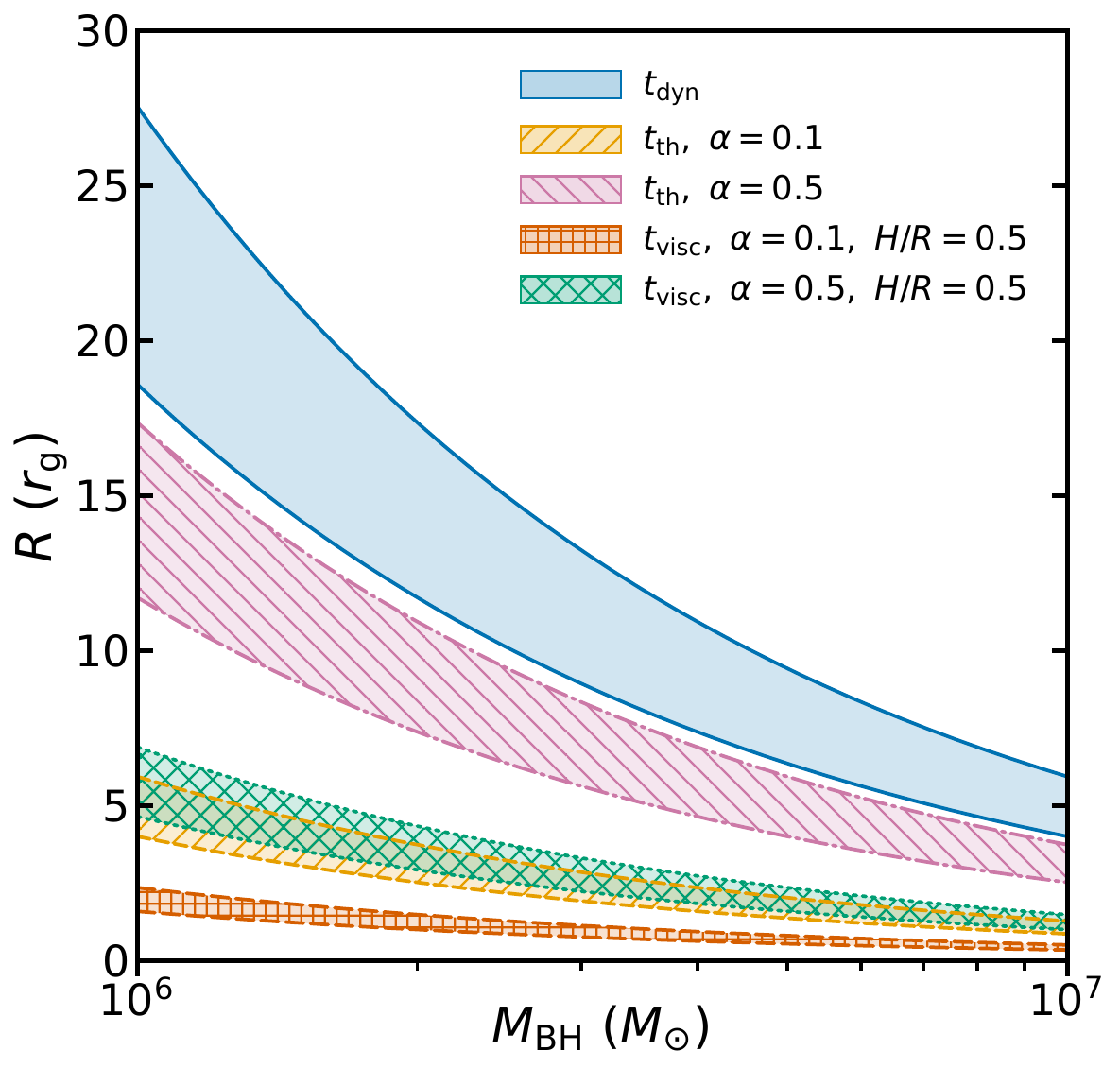}{0.5\textwidth}{}
	}
	\caption{Ranges of the inferred truncation radius as a function of black hole mass under different physical-timescale interpretations of the measured CDT. The blue shaded region corresponds to the dynamical-timescale interpretation, while the orange and pink hatched regions represent the thermal-timescale interpretations for $\alpha=0.1$ and $0.5$, respectively. The red-orange plus-hatched and green cross-hatched regions show the viscous-timescale interpretations for $\alpha=0.1$ and $0.5$, respectively, with $H/R=0.5$ in both cases. For each interpretation, the upper and lower boundaries are derived from the maximum and minimum measured CDTs in the source rest frame, $\ln\tau_{\rm rest}=6.61$ and $6.02$, respectively. The black hole mass is allowed to vary over $10^{6}$--$10^{7}\,M_{\odot}$, and the inferred radii are expressed in units of the gravitational radius.\label{fig:model}}
\end{figure}

\section{Discussion} \label{sec:dis}
\subsection{Possible Physical Origins of the Hard- and Soft-band CDTs}
In this work, we modeled the multi-epoch light curves of RE J1034+396 using the DRW model and obtained 17 reliable CDT measurements. The hard-band CDTs were generally shorter than their soft-band counterparts. Most notably, we report for the first time a QPO-like coevolution between CDTs and X-ray time lags among AGNs. In both energy bands, the CDT and time lag trace closed loops in parameter space that resemble the previously reported QPO frequency--time-lag loop. However, the soft-band loop deviates more strongly from the QPO frequency--time-lag pattern and exhibits a more complex internal structure. Along the hypothetical evolutionary sequence proposed by \citet{2025ApJ...983...13X}, both the hard- and soft-band CDTs exhibit systematic variations, with relative phase offsets with respect to the QPO frequency and X-ray time lag. The presence of these phase offsets accounts for the deviations of the CDT--time-lag loops from the QPO-frequency--time-lag pattern, particularly in the soft band.

The DRW CDT represents the exponential relaxation timescale over which a perturbation is damped, while the process loses memory of its previous state and returns toward its long-term mean. In black hole accretion systems, it is natural to associate this timescale with the dynamical, thermal, and viscous timescales of the accretion flow \citep{2009ApJ...698..895K,2021Sci...373..789B}. These timescales can be expressed as

\begin{equation}
t_{\rm dyn}
=
16.8
\left(
\frac{M_{\rm BH}}{10^{8}M_{\odot}}
\right)
\left(
\frac{R}{200r_{\rm g}}
\right)^{3/2}
{\rm days},
\label{eq:tdyn}
\end{equation}
and
\begin{equation}
t_{\rm th}
\simeq
\alpha^{-1}t_{\rm dyn}
\simeq
\left(\frac{H}{R}\right)^{2}t_{\rm vis},
\label{eq:timescale_relation}
\end{equation}
where $\alpha$ is the dimensionless viscosity parameter, $r_{\rm g}$ is the gravitational radius, $R$ is the radial location in the accretion flow, and $H$ is the vertical thickness of the disk. 

Given that the hard X-ray emission in AGNs is thought to originate from the hot inner accretion flow/corona, the viscosity parameter in this region is typically estimated to be in the range of 0.1--0.5 \citep{1998tbha.conf..148N,2012A&A...545A.115K,2013ApJS..207...17L}, with $H/R\sim0.5$ \citep{2014ARA&A..52..529Y}. Within the truncated-disk framework, $R$ represents the truncation radius separating the two accretion structures, i.e., the outer edge of the inner accretion flow \citep{2011MNRAS.415.2323I}. Before applying these physical timescales to our measurements, cosmological time dilation must also be taken into account, such that $\tau_{\rm rest}=(1+z)^{-1}\tau_{\rm obs}$. RE J1034+396 has a redshift of $z=0.043$ \citep{1995MNRAS.276...20P}, and its central black hole mass is estimated to lie in the range of $10^{6}$--$10^{7}\,M_{\sun}$ \citep{2016A&A...594A.102C}. Using the logarithmic hard-band CDT range of 6.02--6.61 measured from the observations included in Figures~\ref{fig:loop} and \ref{fig:sequence}, we calculated the corresponding ranges of truncation radii under the three different physical-timescale interpretations (see Table~\ref{tab:model}). 

Furthermore, we provided the inferred truncation-radius ranges as a function of black hole mass under the dynamical, thermal, and viscous interpretations of the measured CDTs in the source rest frame (see Figure~\ref{fig:model}). The calculations were performed over the black hole mass range of $10^{6}$--$10^{7}\,M_{\odot}$ using the minimum and maximum CDT values after correcting for cosmological time dilation. Different values of $\alpha$ and $H/R$ were adopted to illustrate how the inferred radius depends on the assumed accretion-flow properties. Given that the X-ray-emitting coronae of AGNs are generally inferred to have outer radii not exceeding about $10\,r_{\rm g}$ \citep{2025FrASS..1130392L}, we adopt this value as an approximate upper limit for the hot inner flow/corona. As shown in Figure~\ref{fig:model}, this upper limit places an additional constraint on the black hole mass under the dynamical-timescale interpretation. The thermal-timescale interpretation with $\alpha=0.5$ also excludes part of the lower end of the adopted mass range. In contrast, the thermal-timescale case with $\alpha=0.1$ and both viscous-timescale cases with $H/R=0.5$ remain below $10\,r_{\rm g}$ throughout the full mass range of $10^{6}$--$10^{7}\,M_{\odot}$ and therefore impose no additional constraint on $M_{\rm BH}$. Although the inferred radii depend on the adopted physical timescale, our results support a dynamic picture in which the radial extent of the hot inner flow/corona evolves with time \citep{2000ApJ...538L.137N,2011MNRAS.415.2323I,2015MNRAS.449..129W,2021MNRAS.503.5522K}.

RE~J1034+396 exhibits a prominent soft X-ray excess, whose physical origin remains under debate. A widely favored interpretation is that the relatively stable underlying soft emission is produced by disk photons undergoing Comptonization in a warm, optically thick plasma covering the disk surface \citep{2009MNRAS.394..250M,2012MNRAS.420.1848D}. Because this warm Comptonizing region may extend to larger characteristic radii than the compact hot corona, variability associated with dynamical, thermal, or viscous processes at these radii is expected to occur on longer timescales \citep{2022ApJ...936...36Y}. The additional smoothing introduced by multiple Compton scattering may further contribute to the systematically longer soft-band CDTs. Within the framework of \citet{2012MNRAS.420.1848D}, the transition radius associated with the warm Comptonizing region may extend to tens of $r_{\rm g}$, further supporting the idea that a substantial fraction of the soft-band emission originates at larger characteristic radii than the hard X-ray emission.

However, the soft-band variability is not determined solely by the intrinsic variability of the warm Comptonizing region. Variations in the hot corona can affect the soft band through both direct and indirect channels. The low-energy extension of the coronal power-law continuum directly introduces coronal variability into the soft band and can account for the observed soft-band QPO, whereas irradiation of the disk and warm Comptonizing plasma produces a delayed soft response \citep{2025ApJ...987..135T}. These two contributions demonstrate that the hot corona directly shapes the observed soft-band timing properties, rather than merely modifying its time-averaged spectrum. They are therefore also likely to make important contributions to the measured soft-band CDT, which should be regarded as an effective timescale resulting from the mixture of intrinsic warm-plasma variability, delayed reprocessing, and direct coronal contamination.

\subsection{Implications of the QPO-like Coevolution for the Inner Accretion Structure}
In RE J1034+396, \citet{2025ApJ...987..135T} provided strong evidence that the QPO originates in the hot corona. They showed that a periastron-precession interpretation would constrain the outer radius of the hot inner flow to approximately $5$--$12\ r_{\rm g}$, whereas a Lense--Thirring precession interpretation would require an outer radius of less than approximately $6\ r_{\rm g}$. Notably, the radial constraints that we infer for the origin of the hard-band CDT variability largely overlap with these ranges (see Table~\ref{tab:model} and Figure~\ref{fig:model}). This consistency and the QPO-like CDT–lag coevolution found in this work indicate a close connection between these two distinct timing properties. Consequently, our result therefore provides the first direct observational support among AGNs for the physical framework proposed by \citet{2011MNRAS.415.2323I}, in which changes in the spatial extent of the hot inner flow/corona simultaneously affect QPO and stochastic variability. 

Previous studies primarily focused on the lag reversals associated with changes in the QPO frequency. In XRBs, the transition from a hard lag to a soft lag has been attributed to systematic changes in the coronal size and disk--corona feedback efficiency as the QPO frequency evolves \citep{2000ApJ...538L.137N,2021MNRAS.503.5522K}. In RE~J1034+396, \citet{2025ApJ...987..135T} interpreted the observed lag reversals as phase wrapping of an intrinsic $\sim2000$~s soft lag as the QPO frequency changes. These studies established a close physical connection between QPO evolution, coronal structure, and the observed X-ray time lag.

Building on this picture, our results further reveal that the QPO frequency, hard-band CDT, and X-ray time lag coevolve with relative phase offsets, tracing counterclockwise closed loops in the corresponding parameter spaces. A plausible interpretation is that these observables trace different aspects of a common cyclic evolution of the inner accretion flow and hot corona. The QPO frequency primarily probes the dynamical state of the QPO-emitting region, whereas the hard-band CDT characterizes the effective relaxation of stochastic coronal variability. The X-ray time lag reflects radiative propagation, disk--corona feedback, and reprocessing. Because these processes may respond to the underlying structural evolution on different timescales, the corresponding observables can acquire relative phase offsets. The closed loops can therefore be understood as different projections of the same multidimensional evolutionary cycle. More broadly, this phase-offset coevolution provides a new observational perspective on the coupled geometric, dynamical, and radiative evolution of the innermost accretion flow and corona in AGNs.
\section{Conclusions} \label{sec:conclusion}
Previous studies of RE~J1034+396 have primarily focused on the evolution of QPO frequency and X-ray time lag, revealing a close connection between the QPO and the physical state of the hot corona. However, the relationship between the QPO and the stochastic variability has remained less well explored. In this work, we investigate the CDTs in the soft and hard X-ray bands across multiple epochs and identify a QPO-like coevolution between the CDT and X-ray time lag. Our main findings can be summarized as follows:

\begin{enumerate}
    \item Across the multi-epoch observations in this source, the soft-band CDT was consistently larger than the hard-band CDT. The hard-band emission is generally thought to originate primarily from the compact hot inner flow/corona, and our timescale-based estimates place its characteristic radial scale within the innermost $\sim10\ R_{\rm g}$ under plausible dynamical, thermal, and viscous interpretations. The longer soft-band CDT may reflect the combined contributions of variability from a more extended warm Comptonizing region, delayed reprocessing, and direct coronal emission, and should therefore be regarded as an effective rather than uniquely local physical timescale.

    \item In the hard-band CDT--time-lag parameter space, for the first time, the measurements trace a counterclockwise closed loop similar to that observed in the QPO-frequency--time-lag evolution. By comparison, the soft-band CDT deviates more strongly from this loop, with additional complex structures appearing within the trajectory. These results further support a close connection between the hard-band CDT and the QPO, whereas the more complex soft-band behavior may indicate that the soft X-ray-emitting region differs from the region in which the QPO originates.

    \item Building on the physical framework originally proposed for XRBs by \citet{2011MNRAS.415.2323I}, our study provides the first direct observational support among AGNs for the dynamical evolution of the corona through the coupled behavior of three distinct timing observables: the QPO frequency, X-ray time lag, and CDT. By extending this framework from stellar-mass to supermassive black hole systems, our results support a common physical picture for the coupled evolution of the inner accretion flow and corona across different black hole mass scales. 
\end{enumerate}

\begin{acknowledgments}
This work is based on observations conducted by XMM-Newton, an ESA science mission with instruments and contributions directly funded by ESA Member States and the USA (NASA). This work was supported by the NSFC grant (12233006).
\end{acknowledgments}

\facilities{XMM (EPIC)}.
\software{\texttt{celerite} \citep{2017AJ....154..220F}, \texttt{emcee} \citep{2013PASP..125..306F}, \texttt{Astropy} \citep{2013A&A...558A..33A}, \texttt{SAS} \citep{2004ASPC..314..759G}, \texttt{Numpy} \citep{2020Natur.585..357H}, and \texttt{Matplotlib} \citep{2007CSE.....9...90H}.}
\clearpage
\appendix
\onecolumngrid
\section{Lomb--Scargle Periodograms of the Standardized Residuals} \label{sec:appendix}
\begin{figure*}[h]
	\figurenum{A1}
	\gridline{\fig{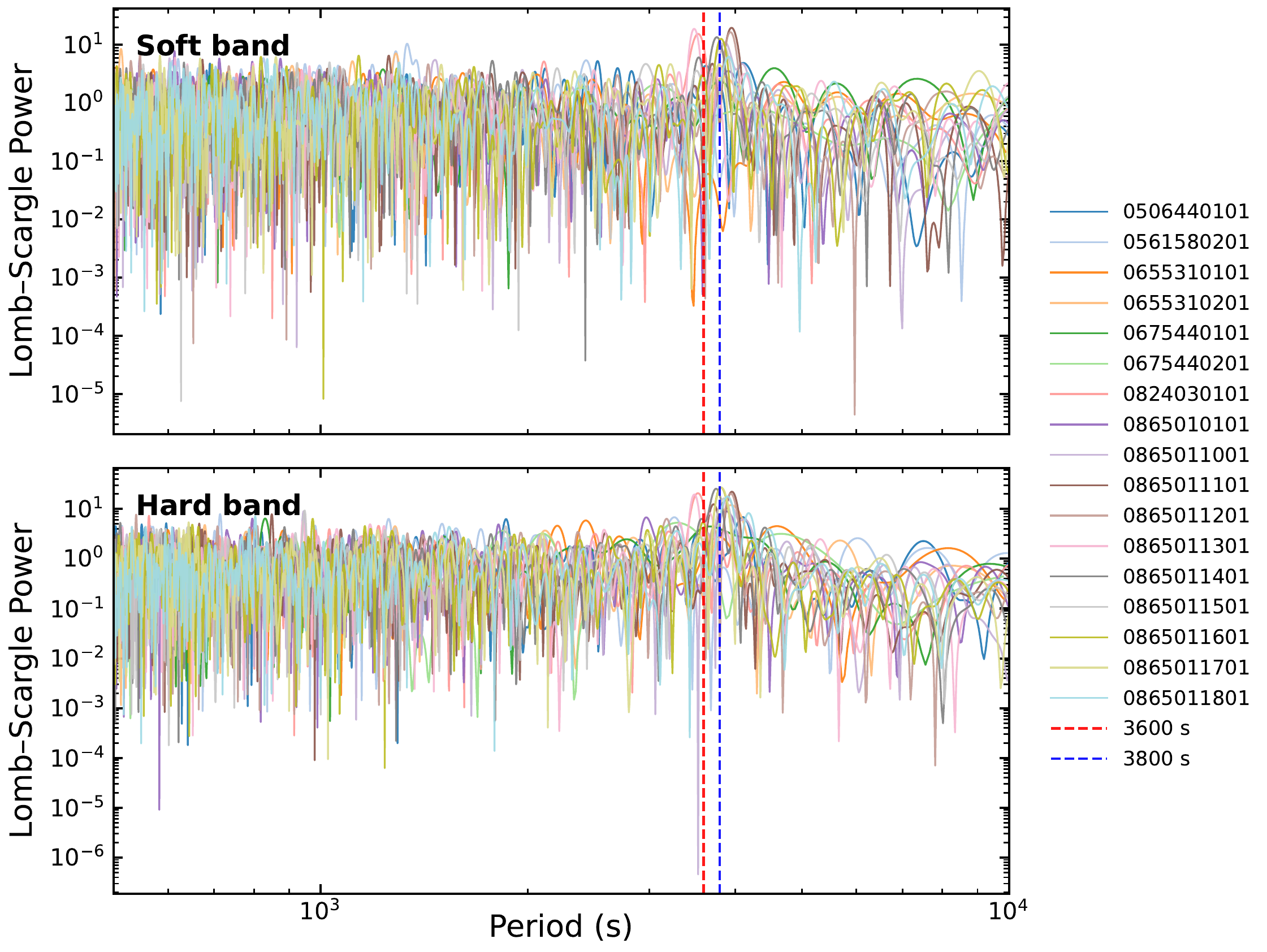}{\textwidth}{}
	}
    \caption{
    Lomb--Scargle periodograms of the standardized residuals obtained from the DRW modeling of the 17 observations that satisfy the parameter-reliability criteria. The upper and lower panels show the results for the soft and hard bands, respectively, with different colors representing different observations. The red and blue vertical dashed lines mark the reference periods of 3600 and 3800~s, respectively, corresponding to the approximate period range of the previously detected QPO. The periodograms are generally consistent with white-noise behavior, although peaks near the QPO period may remain in observations with highly significant QPO detections because the DRW model accounts only for the red-noise component of the light curves.\label{fig:res_lsp}}
\end{figure*}

The Lomb--Scargle periodograms of the standardized residuals from the 17 valid measurements are presented in Figure~\ref{fig:res_lsp}. In both the hard and soft bands, the periodograms are in good agreement with white-noise behavior. Therefore, the DRW model provides a good description of the red-noise variability present in the original light curves.

\section{Robustness of the Loop-like Evolution} \label{sec:appendix2}
To further examine whether the loop-like patterns shown in Figure~\ref{fig:loop} are robust against measurement uncertainties, we calculated the signed areas enclosed by the evolutionary trajectories. Using the parameter uncertainties obtained from the MCMC analysis, we generated $10^{5}$ Monte Carlo realizations of the QPO frequency, X-ray time lag, and hard- and soft-band CDTs. For each realization, the measurements were arranged according to the evolutionary sequence proposed by \citet{2025ApJ...983...13X}, and the corresponding loop area was calculated.

For a sequence of points $(x_i,y_i)$, the signed loop area was calculated using the shoelace formula \citep{Braden01091986},
\begin{equation}
A=\frac{1}{2}\sum_{i=1}^{N}
\left(
x_i y_{i+1}-x_{i+1}y_i
\right),
\end{equation}
where $(x_{N+1},y_{N+1})=(x_1,y_1)$. Before calculating the area, both coordinates were standardized to remove the effects of their different numerical scales. With the adopted axis definitions, positive and negative values of $A$ correspond to counterclockwise and clockwise evolution, respectively.

\begin{figure*}[h]
	\figurenum{A2}
	\gridline{\fig{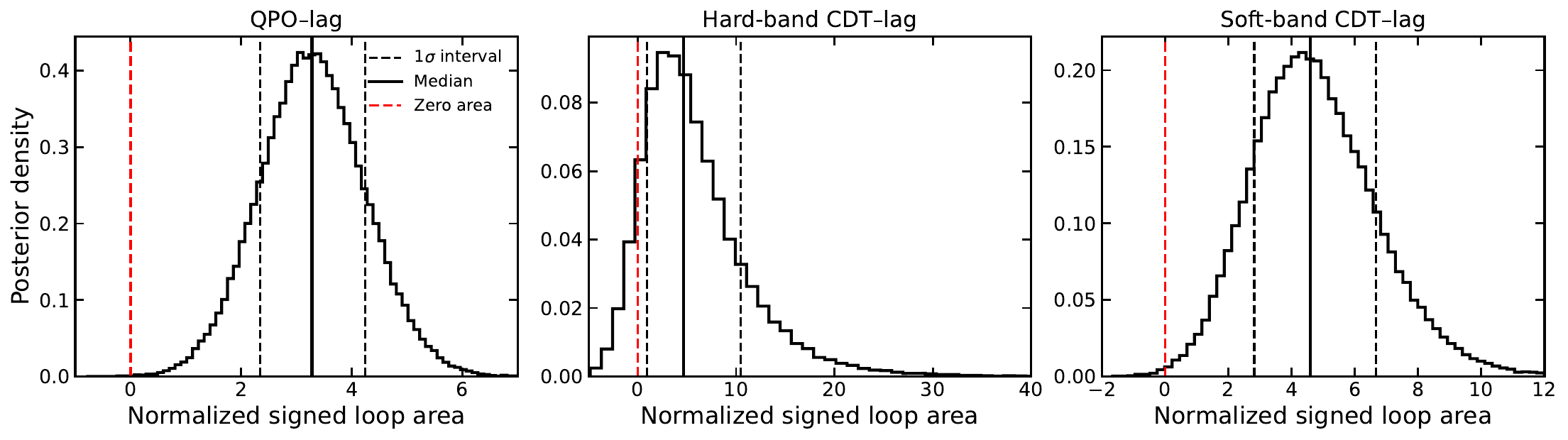}{\textwidth}{}
	}
    \caption{
    Monte Carlo distributions of the normalized signed loop areas for the QPO-frequency--lag (left), hard-band reciprocal-CDT--lag (middle), and soft-band reciprocal-CDT--lag (right) relations along the hypothetical evolutionary sequence proposed by \citet{2025ApJ...983...13X}. The solid black lines indicate the median areas, while the dashed black lines mark the 16th and 84th percentiles, corresponding to the 1$\sigma$ confidence intervals. The red dashed lines denote zero signed area. Positive signed areas correspond to counterclockwise evolution in the respective parameter spaces. All three distributions favor positive signed areas, although the hard-band distribution is substantially broader than those of the QPO and soft-band cases.\label{fig:loops}}
\end{figure*}

Figure~\ref{fig:loops} shows the resulting distributions of the normalized signed loop areas for the QPO-frequency--lag, hard-band reciprocal-CDT--lag, and soft-band reciprocal-CDT--lag relations. The solid black lines indicate the median values, while the dashed black lines mark the 1$\sigma$ confidence intervals. The red dashed lines indicate zero area. The normalized signed areas are $3.28_{-0.93}^{+0.96}$, $4.67_{-3.71}^{+5.79}$, and $4.60_{-1.78}^{+2.07}$ for the QPO-frequency--lag, hard-band reciprocal-CDT--lag, and soft-band reciprocal-CDT--lag relations, respectively. The fractions of Monte Carlo realizations with positive signed areas are 99.975\%, 90.506\%, and 99.746\%, respectively. Thus, the QPO-frequency--lag and soft-band CDT--lag loops remain counterclockwise in almost all realizations. Although the hard-band distribution is considerably broader, it also shows a clear preference for a positive signed area. These results indicate that the observed counterclockwise loop-like evolution is generally robust against the measurement uncertainties.

\newpage
\bibliography{ref}{}
\bibliographystyle{aasjournalv7}
\end{document}